\documentclass[]{aa}
\usepackage{epsfig}
\def\lsim{\vcenter{\hbox{$<$}\offinterlineskip\hbox{$\sim$}}}
\def\gsim{\vcenter{\hbox{$>$}\offinterlineskip\hbox{$\sim$}}}
\begin{document}
\title{Discovery of multiple shells around V838\,Monocerotis}
\author{Jacco Th. van Loon, A. Evans, Mark T. Rushton, Barry Smalley}
\institute{Astrophysics Group, School of Chemistry \& Physics, Keele
           University, Staffordshire ST5 5BG, United Kingdom}
\date{Received date; accepted date}
\titlerunning{Discovery of multiple shells around V838\,Monocerotis}
\authorrunning{van Loon et al.}
\abstract{
We report the discovery of multiple shells around the eruptive variable star
V838\,Mon. Two dust shells are seen in IRAS and MSX images, which themselves
are situated in a shell of CO. This securely establishes V838\,Mon as an
evolved object. We revisit the light echo, which arises from scattering off
the innermost resolved dust shell, to infer a distance to V838\,Mon of
$\gsim5.5$ kpc. The dynamical timescales of the ejected shells, location in
the Milky Way and inferred luminosity are consistent with V838\,Mon being a
low-mass AGB star experiencing thermal pulses of which the 2002 eruption might
have sent the star into the post-AGB phase. This scenario, however, is
inconsistent with the presence of a (young and massive) B3\,V companion.
}
\maketitle

\section{Introduction}

V838\,Monocerotis was discovered on January 6, 2002, (Brown 2002) as a new
eruptive variable star, and peaked at $m_{\rm V}\simeq7$ mag on February 5,
2002. It then declined in the optical whilst becoming brighter in the IR
(Munari et al.\ 2002). This was accompanied by dramatic changes in the
photosphere, making the object an L-type supergiant (Evans et al.\ 2003). The
cause of the eruption of V838\,Mon and the nature of its progenitor are
unclear.

The distance to V838\,Mon may be estimated from the evolution of the light
echo, which was discovered in mid-February 2002 (Henden, Munari \& Schwartz
2002). After first interpretations suggested a distance of $d\sim700$ pc
(Munari et al.\ 2002; Kimeswenger et al.\ 2002), subsequent revisions have led
to ever increasing values for the distance of $\sim2.5$ kpc (Wisniewski et
al.\ 2003), $\gsim6$ kpc (Bond et al.\ 2003) and $8\pm2$ kpc (Tylenda 2004).
Early estimates supported the scenario proposed by Soker \& Tylenda (2003) for
the merger of two main-sequence stars, but later estimates allow for
post-main-sequence scenarios.

We present here evidence for multiple mass-loss events from the progenitor of
V838\,Mon, which have occurred before the 2002 eruption. A dust shell is
visible on IRAS images, and the central object is resolved by MSX. The
double-shell system is situated within a larger shell which is visible on CO
maps. We also revisit the evolution of the light echo and discuss the
implications of our findings for the nature of the progenitor and eruption of
V838\,Mon.

\section{Fossil dust shells seen in infrared emission}

Warm dust, such as found in the vicinity of stars, emits at mid-IR
wavelengths, whilst cold dust, such as encountered in the interstellar medium,
emits at far-IR wavelengths. The sky around V838\,Mon is included in both the
IRAS all sky survey (12, 25, 60 and 100 $\mu$m), and the Galactic plane survey
by the MSX in several mid-IR bands of which band A (8.3 $\mu$m) was the most
sensitive. Both surveys were completed well before the recent outburst of
V838\,Mon (1983 and 1996 for IRAS and MSX, respectively).

\subsection{IRAS data}

We retrieved IRAS images from the IRAS
server\footnote{http://www.astro.rug.nl/IRAS-Server/}, of a
$5^\circ\times5^\circ$ area centred on V838\,Mon (Equ J2000 RA$=7^{\rm
h}04^{\rm m}04^{\rm s}.85$, Dec$=-3^\circ50^\prime51^{\prime\prime}.1$),
corrected for zodiacal light. Because of the scanning, IRAS maps suffer from
striping patterns. The worst of these was corrected by subtracting a
`flatfield', which was created by rotating the image through $-7^\circ$,
median-filtering within a box of $2^\prime$ (12 and 25 $\mu$m) or $4^\prime$
(60 and 100 $\mu$m), averaging the rows, expanding the result to create a
two-dimensional image and rotating through $+7^\circ$.

%
%
\begin{figure}[tb]
\centerline{\psfig{figure=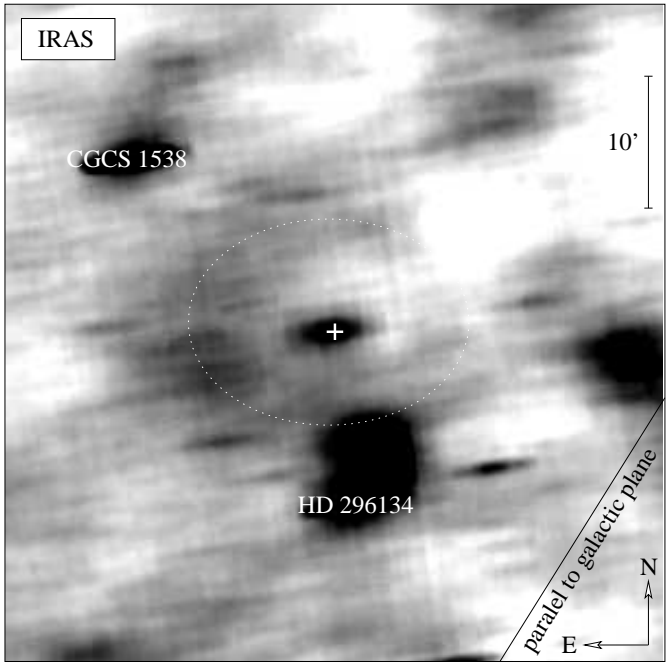,width=88mm}}
\caption[]{IRAS composite image (see text) of a $50^\prime\times50^\prime$
field centred on V838\,Mon (cross). Apart from marginally resolved emission
within $1^\prime$ of V838\,Mon, the object is surrounded by a ring of diffuse
emission (dotted ellipse). The positions of two bright stars are indicated for
guidance.}
\end{figure}

V838\,Mon is clearly detected in all four IRAS bands as a compact source,
surrounded by an elliptical ring of diffuse emission at $\sim7^\prime$ from
the central object in the North-South direction and $\sim10^\prime$ East-West
(Fig.\ 1). The image is an average of the four images, after scaling them to a
similar dynamic range in intensity (this was done by subtracting the mean and
dividing by the mode, which were determined within an area encompassing the
ring). There are two bright IR objects nearby which are identified as the
foreground star HD\,296134 and the dusty carbon star CGCS\,1538, but the
diffuse IR source immediately North of HD\,296134 does not have a known
counterpart.

The presence of a ring-like structure of emission surrounding the position of
V838\,Mon and its persistence in all four IRAS bands lend support to the
significance of its detection. We developed a computer code to test this in a
more objective manner. The procedure is as follows: (1) Along a circle with
radius $r$, the median surface brightness and the difference between the
$3^{\rm rd}$ and $1^{\rm st}$ quartiles of surface brightness values
(``spread'') are determined. The ratio of median and spread is a measure of
the significance of the emission structure. Because a resolved structure is
much more significant than the average signal-to-noise per pixel, the maps are
multiplied by a factor $f=\sqrt{2{\pi}r/\delta}$, where $\delta$ is the size
of a resolution element for which we adopt a conservative $\delta=1^\prime$;
(2) This is repeated taking, in turn, every pixel in the image as the centre
of the circle. Hence a map of the significance of ring-like emission around a
certain position is produced; (3) This is repeated for increasing circle
radii.

This is a conservative approach, as the observed ring structure is clearly
elliptical and a circular ring will therefore never be able to capture all of
the most significant structure. Yet it did recover a ring with a radius of
$10^\prime$ centred less than $1^\prime$ North of the position of V838\,Mon as
the most likely (semi-)circular structure in the image, with a significance of
6.3 $\sigma$.

With an average surface brightness of $S_{100}\sim2$ MJy sr$^{-1}$ the ring
around V838\,Mon is brightest at 100 $\mu$m, where the central source is quite
weak. But with an average surface brightness of $S_{12}\sim0.1$ MJy sr$^{-1}$
it is also clearly visible at 12 $\mu$m where the central source is much
brighter than the ring; probably the dust in the ring is colder than the dust
in the central source. The latter is marginally resolved at 12 and 25 $\mu$m.

\subsection{MSX data}

%
%
\begin{figure}[tb]
\centerline{\psfig{figure=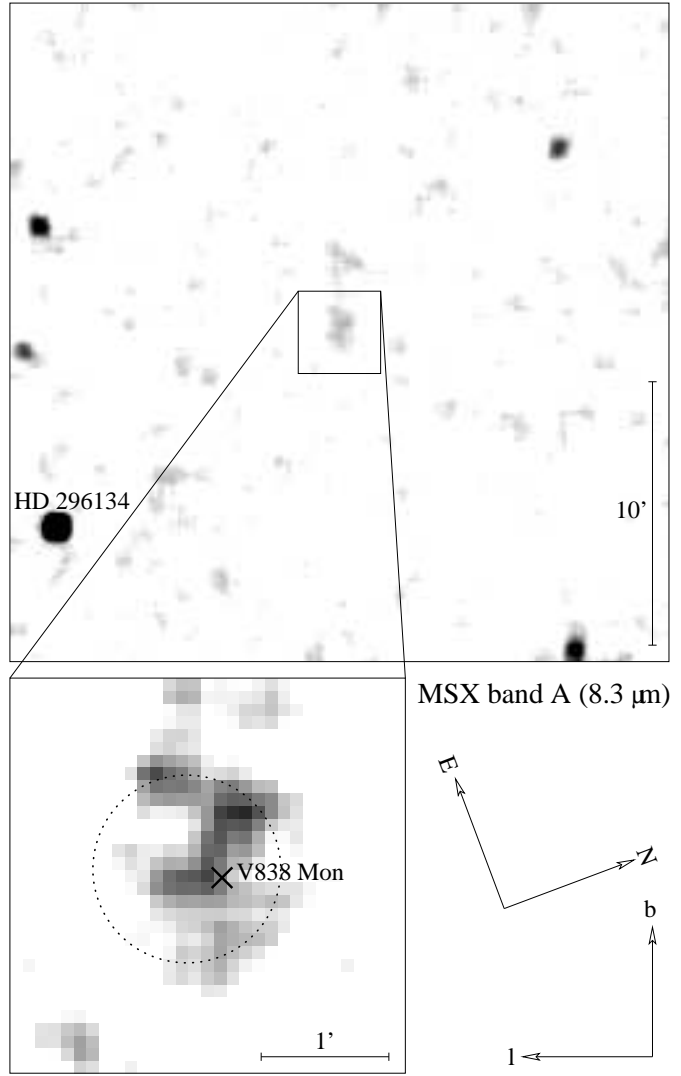,width=88mm}}
\caption[]{MSX band A image (see text) of a $25^\prime\times25^\prime$ field
centred on V838\,Mon, with an enlargement of the central
$3^\prime\times3^\prime$. The position of the bright star HD\,296134 is
indicated for guidance. V838\,Mon is surrounded by emission within a little
less than an arcminute from the central source.}
\end{figure}

The MSX survey reaches a similar depth to the IRAS survey, but the spatial
information within the MSX data is superior. The central source in the IRAS
image is clearly resolved in the MSX data into several emission structures
within a roughly circular area of $\sim1^\prime.5$ diameter (Fig.\ 2). The
band A (8.3 $\mu$m) image was corrected for striping along the Galactic
longitude direction by subtracting an image created from an average of the
columns in between the bright point sources seen in the picture. V838\,Mon is
located near the centroid of the emission complex. None of the emission knots
near V838\,Mon could be identified with stars in 2MASS JHK images (which do
show what is believed to be the progenitor of V838\,Mon).

The significance of the emission is estimated by rebinning the image such that
the emission extends over only two (adjacent) pixels, each with a surface
brightness of $S_{8.3}\simeq2.0\times10^{-7}$ W m$^{-2}$ sr$^{-1}$. Excluding
the emission from the five obvious stars, the standard deviation of the
fluctuations in the rebinned map is $\sigma\simeq5\times10^{-8}$ W m$^{-2}$
sr$^{-1}$; the emission associated with V838\,Mon is significant at the
$\sqrt{2}\times4\sigma$ level, i.e.\ more than 6 $\sigma$. Being the brightest
square arcminute in the entire $25^\prime\times25^\prime$ field (apart from
the five bright stars), and its disposition with respect to V838\,Mon, lead us
to conclude that it is very unlikely to be a chance superposition of
fluctuations in the background. At nearly $4\sigma$, the next-significant
patch of emission in this part of the MSX band A image is the smudge at
$\sim5^\prime$ North of HD\,296134, which may be identified with the bright IR
source above HD\,296134 in the IRAS image (Fig.\ 1).

\section{A large shell of carbon monoxide?}

Due to its proximity to the Galactic plane ($l=217^\circ.80$, $b=+1^\circ.05$)
the sky around V838\,Mon was included in several surveys of emission from
carbon monoxide (CO), of which the latest compilation is presented by Dame,
Hartmann \& Thaddeus (2001). Located in the direction just shy of what looks
like a molecular cloud, V838\,Mon seems to be situated off-centre within a
bubble of CO emission with a diameter of $\sim1^\circ$ (Fig.\ 3).

The eastern rim of the bubble (``upper left'' in Fig.\ 3) has only been
covered in the Superbeam survey (Dame et al.\ 1987), at a relatively poor
spatial resolution ($\Delta\theta=0.5^\circ$). Its velocity-integrated flux
level is $\sim0.1$ K km s$^{-1}$ per angular resolution element. The western
part adjoining the molecular cloud was recorded at a higher spatial resolution
($\Delta\theta=0.25^\circ$; May et al.\ 1993), and is also discernible on 60
and 100 $\mu$m IRAS maps. The flux level of the edge of the molecular cloud
delineating this part of the bubble is $\sim2$ K km s$^{-1}$ per angular
resolution element. We note that in constructing the composite CO map, Dame et
al.\ only integrated emission that exceeded $3\times$ the rms noise level in
the spectra, and hence any emission seen in their maps is statistically
significant.

The radial velocity of V838\,Mon of $v_{\rm LSR}\simeq+50$ km s$^{-1}$ (Kipper
et al.\ 2004) is rather high and suggests a location in the outskirts of the
Milky Way disc (Fig.\ 3 in Dame et al.\ 2001). Given the fact that there is no
other bubble to be seen within an area of {\it at least} 10 square degree, the
chance for the line-of-sight towards V838\,Mon to intersect the bubble (which
covers an area less than 1 square degree) by sheer coincidence is certainly
$<10\%$. New, sensitive high-spatial resolution CO (and radio continuum) maps
are desired in order to confirm (or otherwise) the physical association of the
CO bubble with V838\,Mon.

%
%
\begin{figure}[tb]
\centerline{\psfig{figure=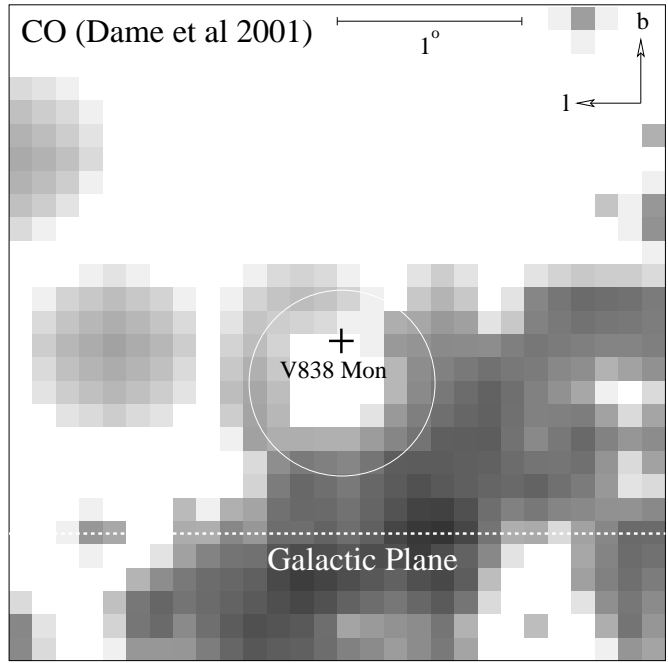,width=88mm}}
\caption[]{A closeup of the Galactic CO map of Dame, Hartmann \& Thaddeus
(2001), centred on V838\,Mon which seems to be situated within a shell of CO
emission $\sim1^\circ$ in diameter.}
\end{figure}

\section{The light echo revisited}

Since its discovery in February 2002, the light echo of V838\,Mon's eruption
has been monitored and used to infer the distance to V838\,Mon. The latest
interpretation was presented by Tylenda (2004), who analysed images of the
light echo taken in 2002. However, images are already available over a much
larger timespan, and in light of our discovery of circumstellar dust shells
around V838\,Mon it is timely to revisit how the light echo can inform us
about the distance and nature of V838\,Mon.

%
%
\begin{table}
\caption[]{Light echo measurements from images spanning the period March 2002
until February 2004. The times are with reference to $t_0 = 5$ February 2002
(brightest visual maximum).}
\begin{tabular}{lrrrr}
\hline\hline
Observatory                                      &
$t - t_0$                                        &
\multicolumn{2}{c}{diameter ($^{\prime\prime}$)} &
ratio                                            \\
                                                 &
(days)                                           &
East-West                                        &
North-South                                      &
                                                 \\
\hline
WHT$^1$  &  53 & 12.0 & 11.2 & 1.07 \\
HST$^2$  &  86 & 14.3 & 14.0 & 1.02 \\
SAAO$^3$ & 101 & 16.4 & 15.3 & 1.07 \\
HST$^2$  & 106 & 17.2 & 16.2 & 1.06 \\
HST$^2$  & 211 & 24.4 & 23.1 & 1.06 \\
SAAO$^3$ & 224 & 24.2 & 23.1 & 1.05 \\
HST$^2$  & 267 & 27.5 & 25.3 & 1.09 \\
USNO$^4$ & 291 & 29.3 & 27.0 & 1.09 \\
HST$^2$  & 317 & 29.7 & 27.9 & 1.06 \\
USNO$^4$ & 625 & 42.8 & 40.2 & 1.06 \\
HST$^2$  & 735 & 46.5 & 43.6 & 1.06 \\
\hline
\end{tabular}
{\small
$^1$Munari et al.\ 2002\\
$^2$http://hubblesite.org/newscenter/newsdesk/archive/releases/ \hspace*{0.5mm}
2003/10 and 2004/10\\
$^3$http://da.saao.ac.za/news/v838mon.jpg\\
$^4$http://www.ghg.net/akelly/v838lar2.jpg and v838lar3.jpg
}
\end{table}

We measured the diameter of the outer rim of the light echo visible on high
quality images taken at the WHT, HST, SAAO and USNO over a period of more than
two years since outburst (Table 1). There are many sources of uncertainty
entering the error on the measured value, but the most important of these ---
irregularity and limited contrast of the rim --- are small enough for a
sensible analysis, and the values given are accurate to $\pm1^{\prime\prime}$.
Although the echo is clearly elliptical, the axial ratio (EW over NS) of
$1.063\pm0.006$ is quite close to unity.

%
%
\begin{figure}[tb]
\centerline{\psfig{figure=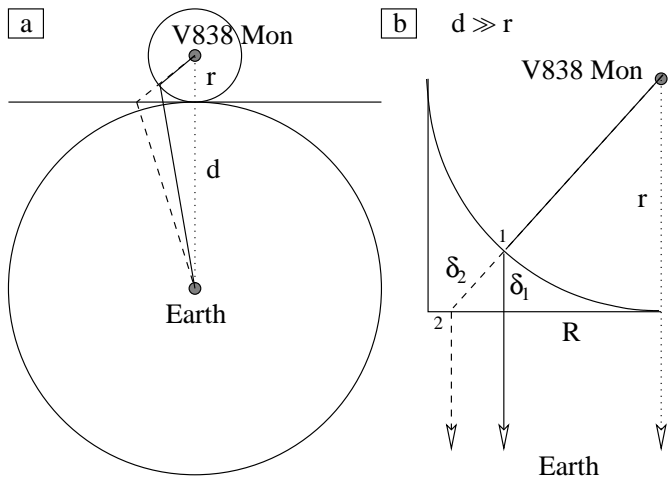,width=88mm}}
\caption[]{Geometry of the light echo effect when (1) scattered off a
circumstellar shell with radius $r$ (solid line), or (2) scattered off a sheet
of interstellar material at distance $r$ in front of V838\,Mon (dashed). The
time delay ${\Delta}t$ with respect to the light traveling directly towards
Earth (dotted) is related to the differential pathlength $\delta=c {\Delta}t$.
For the case $d{\gg}r$ (panel {\bf b}) the pathlength (resp.\ $\delta_1$ or
$\delta_2$) has a simple dependence on $r$ and projected distance $R$, with
$\tan{\theta}{\simeq}R/d$.}
\end{figure}

Simple models for the evolution of the light echo, valid for the small
angular extent of the echo, predict that the diameter grows with time since
outburst, $\Delta t$, as:
\begin{equation}
{\rm diameter} = 2 \arctan{\frac{\sqrt{c \Delta t (2r \pm c \Delta t)}}{d}}
\end{equation}
(see Fig.\ 4) where the minus sign is appropriate for a thin shell of radius
$r$ and the plus sign for a thin plane of scattering dust at a distance $r$ in
front of the light source and perpendicular to the line-of-sight ($c$ is the
speed of light and $d$ is the distance to the light source).

Both the optical images of the light echo and the infrared emission as imaged
by MSX show significant substructure in the scattering medium. Nevertheless,
the evolution of the light echo diameter is remarkably monotonic in time
(Fig.\ 5), which gives credence to the employment of simple models to describe
the global geometry of the scattering medium.

%
%
\begin{figure}[tb]
\centerline{\psfig{figure=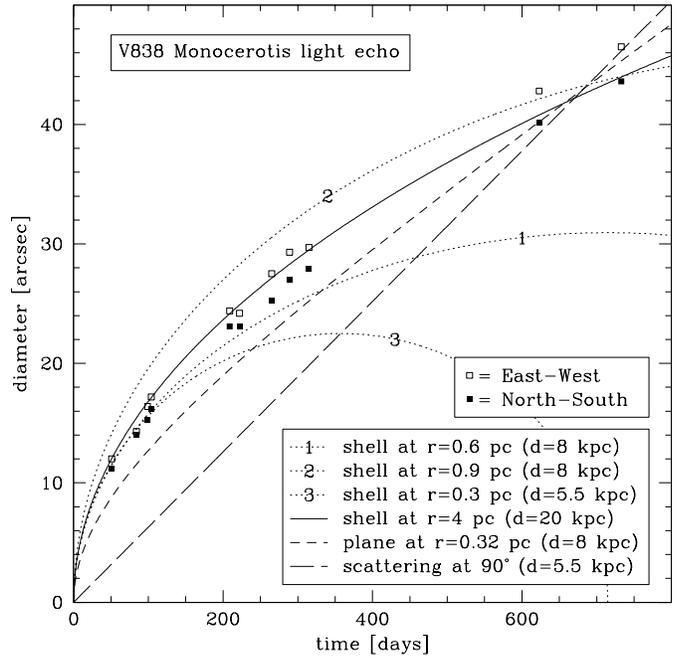,width=88mm}}
\caption[]{Measured evolution of the light echo around V838\,Mon, and the
predictions for several simple models (Equation 1).}
\end{figure}

At a distance $d=8$ kpc, a small shell with radius $r=0.6$ pc (projected
radius $0.^\prime$26) fits the early 2002 data but is inconsistent with the
more recent data, whereas a larger shell with radius $r=0.9$ pc (projected
radius $0.^\prime39$) fits the 2004 data but not the earlier data (Fig.\ 5).
The only single shell solution is obtained by placing V838\,Mon at $d\gsim20$
kpc, allowing for a larger shell (projected radius $>0.^\prime7$). This would
place V838\,Mon beyond the edge of the Milky Way disc, either within the Milky
Way halo or within an as yet unrecognised dwarf galaxy. This is extremely
unlikely, given the proximity of V838\,Mon to the Galactic plane and its
apparent association with a CO shell at the edge of a molecular cloud.

Alternative solutions for $d\sim8$ kpc include a hybrid solution of a thin
shell (with radius $r\sim0.7$ pc) and a thin plane, but the plane would have
to be only $r\sim0.3$ pc in front of V838\,Mon and hence intersect the shell
--- an unlikely and rather {\it ad hoc} geometry.

The data may be reconciled with a reasonable distance if we relax the
assumption of a thin scattering medium. For a distance of $d=5.5$ kpc, the
most recent data may be reproduced by scattering under right angles ($r=0$ in
Eq.\ 1) from dust at $\sim0.6^\prime$ from V838\,Mon. Also at $d=5.5$ kpc, the
early 2002 data may be reproduced by scattering off a shell with $r\sim0.3$ pc
(projected radius $0.19^\prime$). The rim of the light echo might first arise
from scattering off the dense inner edge of the shell, but continue to advance
through the shell outward after the reflection off the inner shell edge turns
back into itself after about a year. The circumstellar material echoing the
light of the February 2002 outburst is thus identified with the innermost
resolved dust shell as seen in the MSX image. For small dust grains ($\lsim0.1
\mu$m) scattering under angles greatly different from $0^\circ$ is
inefficient; the outer rim of the echo will dim faster than by geometric
dilution alone.

This scenario will not work for a distance less than $d\sim5.5$ kpc, because
once scattering occurs under right angles we expect the projected expansion of
the light echo to proceed linearly with time --- and this is not yet seen. But
it works for larger distances, where the scattering does not yet occur under
right angles. We thus predict the light echo to enter a stage of linear
expansion with time, which then yields an unambiguous value for the distance.

\section{The nature of V838\,Monocerotis}

Despite a geometric method (the light echo) to estimate the distance to
V838\,Mon there remains uncertainty about its value as it depends on our
interpretation of the circumstellar dust environment and the scattering
properties of the dust grains. Although the simplest model for the light echo
would place V838\,Mon outside the Milky Way, we also present an interpretation
where the echo proceeds through an extended shell, whereby the scattering
angle approaches $90^\circ$. Accepting the latter as the more likely scenario,
we adopt a distance of $d=5.5$ kpc, but remind the reader that this is a lower
limit.

To determine the time at which the fossil shells were ejected we also need to
know the expansion velocity. Although this might, in principle, be estimated
for the inner shell from spectra of the scattered light echo, such an
observation has yet to be made and will not be feasible for the outer dust
shell. The CO shell expansion velocity might be measured from spectra of the
CO line profile in the line-of-sight towards V838\,Mon, but this has not yet
been feasible (cf.\ Rushton et al.\ 2003) due to the great amount of dilution
rendering only the rim visible (where the radial velocity due to expansion is
minimal).

We here calculate dynamical timescales by {\it assuming} a constant outflow
velocity of $v=10$ km s$^{-1}$. This would yield $10^5$ yr per projected
parsec. A further complication in case of the CO shell is that V838\,Mon is
placed considerably off-centre and that the material might have slowed down as
it had to plough through surrounding interstellar medium. For the subsequent
dust shells this may not have been an issue as the CO shell would have swept
clean a cavity for the dust shells to expand freely into, unless the removed
interstellar medium was replenished by stellar mass loss in between eruptions.

We thus estimate dynamical timescales, for the CO shell $t_1>5\times10^6$ yr,
for the outer dust ring $t_2\sim1.4\times10^6$ yr, and for the inner dust
ring $t_3\sim1.1\times10^5$ yr. If the current eruption is the fourth such
event, then it seems that these events have occurred at an increasing rate.

Placing V838\,Mon at a distance of (at least) 5.5 kpc makes its outburst peak
at a bolometric luminosity $L>10^5$ L$_\odot$. The total energy radiated
throughout the eruption (Bond et al.\ 2003) must then have been $E>10^{38}$ J.
The putative progenitor would have had $L\gsim10^2$ L$_\odot$. The 2002
eruption and possibly earlier eruptions that produced the fossil shells might
thus have been the result of helium-flashes (thermal pulses) on the AGB.
Indeed, Kipper et al.\ (2004) report enhanced abundances of s-process
elements, suggesting $3^{\rm rd}$-dredge-up on the thermal-pulsing AGB.

Situated in the outskirts of the Galaxy, V838\,Mon is likely to be metal poor
(Kipper et al.\ (2004) estimate $[{\rm Fe/H}]\simeq-0.4$) and of low mass
($M\lsim1$ M$_\odot$). Inter-pulse timescales are longer for lower-mass AGB
stars, typically several $10^5$ yr for a 1 M$_\odot$ star, but decrease in
time as the core mass grows (Vassiliadis \& Wood 1993), in accordance with the
above dynamical timescales of the shells around V838\,Mon. An AGB star with
little envelope mass will be severely affected by a thermal pulse at the base
of its envelope. We therefore propose that V838\,Mon experienced its fourth
and final thermal pulse, resulting in the ejection of what may become a
planetary nebula.

Our scenario for a low-mass AGB star progenitor of V838\,Mon, however, is
inconsistent with the presence of a B3\,V companion (Munari, Desidera \&
Henden 2002). Membership of a binary system in which the companion is a young
and massive star implies that the erupting star must be young and massive too.
Future studies must find a way of reconciling these two contrasting sets of
observational evidence.

\begin{acknowledgements}
We would like to thank Dr.\ Joana Oliveira for valuable discussions, and an
anonymous referee for her/his critical remarks which helped improve the
presentation of the results. This research made use of the IRAS data base
server of the Space Research Organisation of the Netherlands (SRON), data
products from the Midcourse Space Experiment, the NASA/ IPAC Infrared Science
Archive, and Aladin. MTR is supported by a PPARC studentship.
\end{acknowledgements}

\end{document}